\documentclass[11pt,twoside]{article}

\usepackage{asp2006}
\usepackage{epsf}
\usepackage{lscape}
\usepackage{graphicx}
\usepackage{epstopdf}
\usepackage{natbib}

\begin{document}
\title{Distinguishing between AGN and Star-forming Galaxies in ATLAS} 
\author{Kate E. Randall,$^{1,2}$ Andrew M. Hopkins,$^{1,3}$, Ray P. Norris,$^2$ and Minnie Y. Mao$^{4,2,3}$} 
\affil{
{\itshape $^1$Sydney Institute for Astronomy, School of Physics, University of Sydney, NSW, 2006, Australia}\\
{\itshape $^2$CSIRO Australia Telescope National Facility, PO Box 76, Epping, NSW, 1710, Australia}\\
{\itshape $^3$ Anglo-Australian Observatory, PO Box 296, Epping, NSW, 1710, Australia}\\
{\itshape $^4$ School of Mathematics and Physics, University of Tasmania, Private Bag 37, Hobart, 7001, Australia}\\}

\begin{abstract} 
The Australia Telescope Large Area Survey (ATLAS) is the widest deep radio survey ever attempted,
covering 7 square degrees of sky in two separate fields, with extensive multi-wavelength data.
The primary aim of this research is to investigate all possible discriminants between active galactic
nuclei (AGN) and star-formation (SF) in ATLAS, with the goal of comparing discriminants, identifying
the strengths and weaknesses, and establishing an optimum technique given the available data.
Ultimately, all possible discriminants will be utilized, including optical$/$infrared SEDs, spectroscopic
line widths, optical line ratios, radio spectral indices, variability, morphology, polarization and the
radio$/$FIR correlation. A preliminary investigation using only the available spectroscopic data in
ATLAS is ongoing. Results from this investigation are presented, exploring the proportion of
AGN$/$SF galaxies as a function of radio flux density down to 150$\mu$Jy. Three faint GPS candidates are also presented, as a preliminary result from ATLAS. 
\end{abstract}

\section{Introduction} 
The Australia Telescope Large Area Survey (ATLAS) \cite{enno,atlas} is the widest deep radio survey so far, covering approximately seven square degrees over two fields; the European Large Area ISO Survey South 1 Region (ELAIS-S1) and the Chandra Deep Field South (CDFS). The current sensitivity of the survey is $\approx\,$30$\mu$Jy, with the aim to achieve an rms of $\approx$10-15$\mu$Jy. These two fields in particular were chosen as they both have deep IR data, from the SWIRE project \cite{lonsdale}, will have deep optical data and X-ray data.
ATLAS has seven primary science goals:
\begin{enumerate}
\vspace{-3mm}\item What is the evolutionary relationship between AGN and star-forming galaxies (SFGs)? \citep[See][]{mao}
\vspace{-3mm}\item Are there any high redshift AGNs in ATLAS? I.e. What are the highest redshift AGNs to be found? What are their properties?
\vspace{-3mm}\item Does the radio-far infrared (FIR) correlation change, depending on the galaxy properties, or with redshift?
\vspace{-3mm}\item How does the mass dependent star formation rate evolve with time?
\vspace{-3mm}\item Can the radio luminosity function be traced to high redshift?
\vspace{-3mm}\item Can the cosmic magnetic field strength be measure with ATLAS?
\vspace{-3mm}\item Are there rare objects that can only be discovered with a wide deep survey such as ATLAS? 
\end{enumerate}
\vspace{-2mm}
ATLAS is a continuing project, with scheduled observations using both the Australia Telescope Compact Array (ATCA) and AAOmega on the Anglo-Australian Telescope (AAT). Already within ATLAS there are several interesting results, including objects known as ``Buried AGN" \cite{buried}, that have the optical properties of a star-forming galaxy, but which reveals an AGN at radio wavelengths. This highlights the importance of being able to distinguish between an AGN and an SFG robustly, and determine the relationship between the two. The distinction between AGN and star-forming galaxies depends on the primary source of energy output from the object. 

\subsection{How to distinguish an SFG from an AGN}
\begin{figure}
\begin{minipage}{0.5\textwidth}
\includegraphics[scale=0.4]{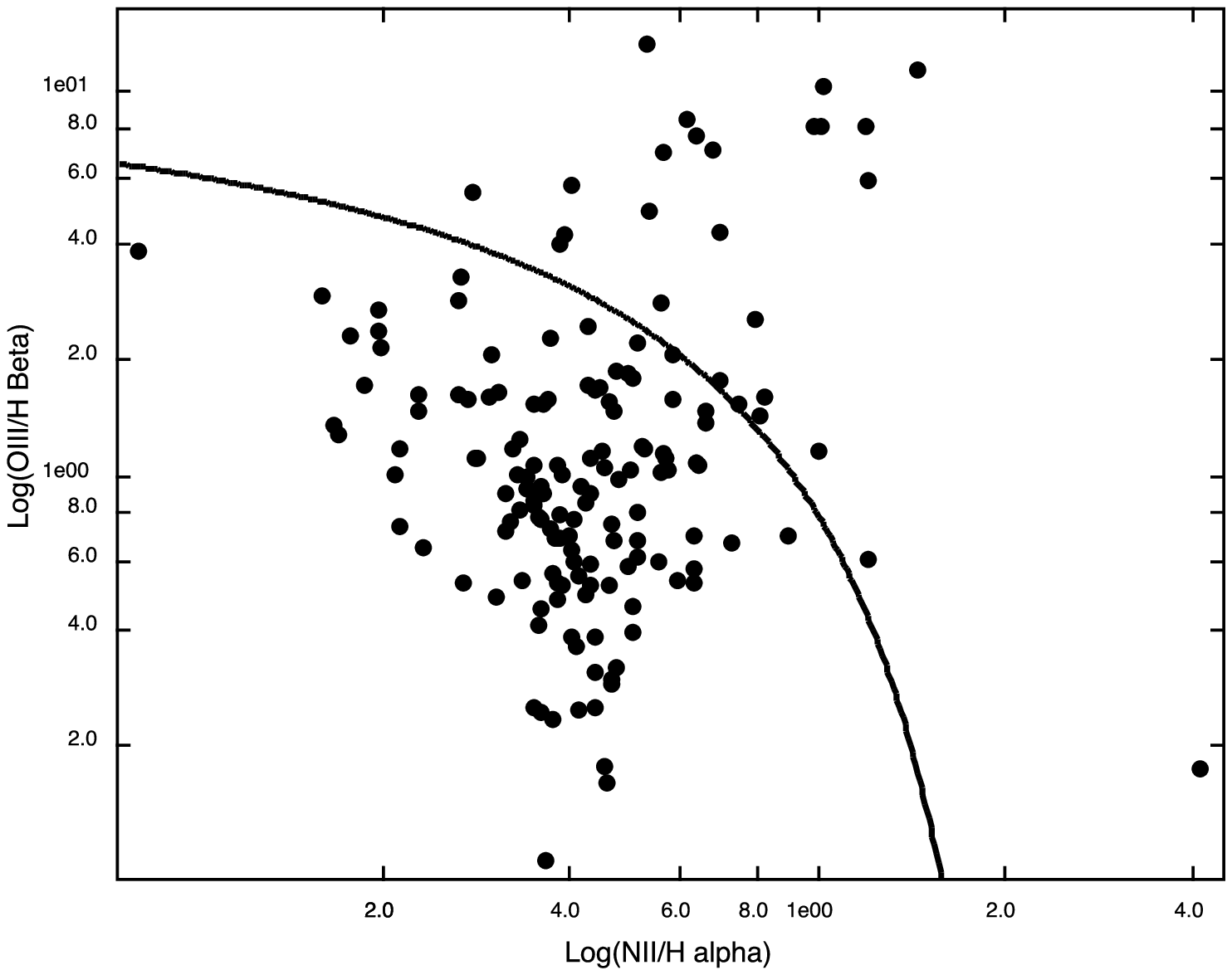}
\caption{Optical line ratio diagram (Baldwin, Phillips \& Terlevich 1981), for
ATLAS data, containing 154 sources, all from CDFS. The boundary line between AGN and SFGs is from \citealt{kewley}.}
\label{fig:bpt}
\end{minipage}
\hspace{0mm}
\begin{minipage}{0.5\textwidth}
\includegraphics[scale=0.6]{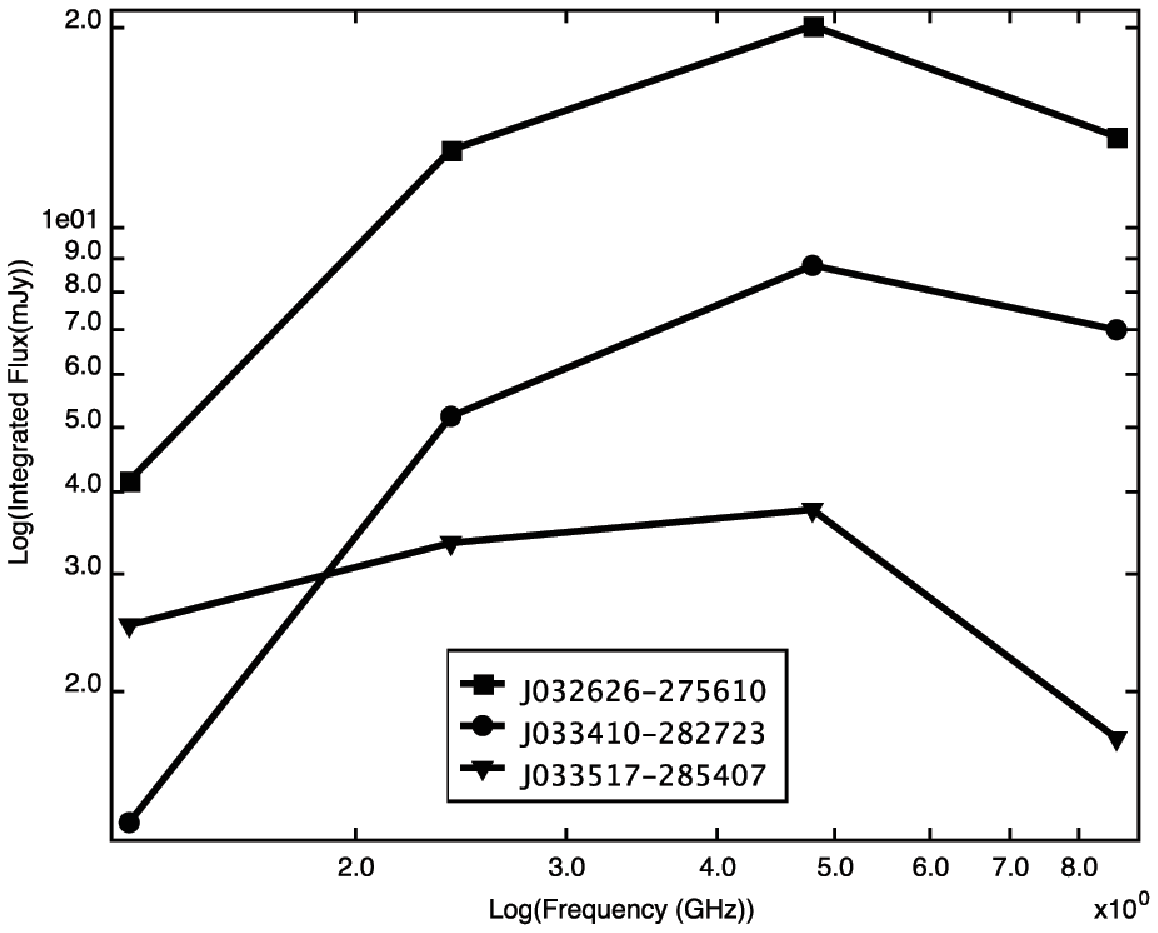}
\caption{The three faint GPS candidates in ATLAS}
\label{fig:gps}
\end{minipage}
\end{figure}

All possible discriminants for distinguishing between and AGN and star-forming galaxy will be utilized for this purpose, including indicators such as optical/infrared spectral energy distributions (SEDs), spectroscopic line widths, Lyman drop-outs, optical line ratios, radio spectral indices, morphology, polarization, variability, VLBI and the radio-FIR correlation. No previous work has systematically compared and combined all these indicators, and hence, upon completion of ATLAS, we intend to present a robust method to distinguish AGN from SFGs. 
Preliminary results of optical line ratios for CDFS from the current spectroscopic data have been completed, as shown in Figure~\ref{fig:bpt}. Thus far, spectroscopy is available for 564 of the ATLAS sources (mostly the brighter objects), both from the literature and utilizing AAOmega on the AAT  \citep[for more detailed information see][]{mao}. To gain spectroscopy for the fainter ATLAS sources, proposals will be submitted to larger optical telescopes, such as Gemini. 

A preliminary comparison of the position of sources on the optical line ratio diagram with their position on the radio-24$\mu$m correlation has been completed. As a result of this analysis, it was discovered that the AGN in the upper right hand corner of Figure~\ref{fig:bpt} (i.e., the strong AGN detected in ATLAS), had amongst the highest radio-24$\mu$m ratios. A more thorough investigation of these sources is in progress, to understand why this occurs and what consequences it will have upon our understanding of AGN. This is possibly another example of an unusual result from a wide deep survey such as ATLAS, as previous comparisons of AGN$/$SFG indicators have not been so extensive.  

\section{Faint Gigahertz Peaked Spectrum (GPS) Sources in ATLAS}
Gigahertz Peaked Spectrum (GPS) sources are also of interest within ATLAS, due to their probable links to AGN and galaxy evolution \cite{odea}. Due to their similar morphology and power, it is thought that GPS sources evolve into large scale radio sources such as FR-I/II galaxies. GPS sources constitute approximately 10$\%$ of the bright radio source population. If there is a similar fraction of GPS sources at faint flux densities, we would expect several hundred of the $\approx$2000 radio sources currently catalogued in ATLAS to fall into this category.  In June, 2008, we utilized ATCA to observe six possible GPS candidates within ATLAS. Three of these candidates showed the characteristic spectral turnover of GPS sources, as shown in Figure~\ref{fig:gps}. 

\section{Future work}
The extensive multi-wavelength data from ATLAS will allow an in-depth investigation into the star formation history of the Universe and give insights into the evolution of radio sources, including AGN and SFGs, GPS and CSS sources. The sources described in this paper, such as the extreme AGN, will continue to be investigated and we will also search for more possible candidates for GPS and CSS sources within ATLAS data. Extra radio data will be added to the ATLAS catalogues, from telescopes such as Molonglo Observatory Synthesis Telescope (MOST) and the Giant Metre Radio Telescope (GMRT), aiding in this investigation. A more comprehensive optical line ratio diagram will be completed as more spectroscopic data is obtained, which will need to be from larger optical telescopes than the AAT, such as Gemini, to gain data for the fainter ATLAS sources. 



\end{document}